\documentclass{article}
\usepackage[utf8]{inputenc}
\usepackage{amsmath}
\usepackage{amssymb}

\renewcommand{\>}{\rangle}

\renewcommand{\u}{|\uparrow\>}
\renewcommand{\d}{|\downarrow\>}
\newcommand{\1}{|1\>}
\newcommand{\0}{|0\>}
\newcommand{\x}{\otimes}
\renewcommand{\b}{\Big}

\newcounter{wn}
\newcommand{\WallComp}{\noindent\textbf{Feature (\stepcounter{wn}\roman{wn}):} }

\usepackage{natbib}
\bibpunct{(}{)}{;}{a}{}{,}

\title{Everettian Branching in the World and of the World\footnote{We thank Ruward Mulder, Simon Saunders, Chris Timpson, David Wallace, and two anonymous referees for helpful feedback that suffuses the following. No endorsement of the following is implied by their generosity. Contact: huggett@uic.edu}}

\author{Nadia Blackshaw, University of Bristol\\Nick Huggett, University of Illinois Chicago\\James Ladyman, University of Bristol}

\begin{document}

\maketitle

\begin{quote}
    “There are limitless futures stretching out in every direction from this moment—and from this moment and from this. Billions of them, bifurcating every instant! Every possible position of every possible electron balloons out into billions of probabilities!”

\textit{Mostly Harmless}, Douglas \citeauthor{adams2009mostly}
\end{quote}

\begin{abstract}
    This paper investigates the formation and propagation of wavefunction `branches' through the process of entanglement with the environment. While this process is a consequence of unitary dynamics, and hence significant to many if not all approaches to quantum theory, it plays a central role in many recent articulations of the Everett or `many worlds' interpretation. A highly idealized model of a locally interacting system and environment is described, and investigated in several situations in which branching occurs, including those involving Bell inequality violating correlations; we illustrate how any non-locality is compatible with the locality of the dynamics. Although branching is particularly important for many worlds quantum theory, we take a neutral stance here, simply tracing out the consequences of a unitary dynamics. The overall goals are to provide a simple concrete realization of the quantum physics of branch formation, and especially to emphasise the compatibility of branching with relativity; the paper is intended to illuminate matters both for foundational work, and for the application of quantum theory to non-isolated systems.
\end{abstract}

\section{Introduction}\label{sec:intro}

The Everett or `many worlds' interpretation of quantum theory (QT) can be predicated on denying wavefunction collapse as a physical process, and taking at face value the entanglement that results from unitary time evolution governing the interaction of a quantum system with  a measuring device (or an observer, or anything else including its environment, all treated as further quantum systems). After such interactions, each system has a different `relative state' in with respect to the others, but neither system has a unique state, so reality is branched into `many worlds'. Everett interpretations must thus give an account of how branching works.\footnote{Of course there is also a `probability problem' for Everett, which is not addressed here. See, e.g., \citet{saunders2010MW} for a detailed discussion.} To define the branching structure at an instant requires a choice of basis, but there are also the questions of how the world splits into different branches over time, and whether it is supposed to do so globally or whether branching happens regionally, so that not all the world branches at once.\footnote{There has been some discussion of whether a preferred basis must be specified prior to the branch dynamics (see \citet{barrettquantum1999}, \citet{cunninghambranch2014}). The model here focuses solely on the dynamics by fiat of specifying a basis.} Following the pioneering work of \citet{zurekEnviron1982,zurek2003decoherence},  \citet{saunders1993decoherence}, and \citet{PhysRevD.47.3345}, many Everettians take the branching structure to be determined by the physical process of decoherence, the theory of which is used to select a dynamically stable basis with respect to which the reduced density matrix of a subsystem interacting with the environment is approximately diagonalised. Decoherence is understood to happen according to the normal local -- meaning no action at a distance or superluminal causality -- interactions of quantum systems with each other as described by unitary time evolution. 

The most conceptually developed account of such Everettian branching by decoherence is that of \citet{wallace2012emergent}. It has many different features, which interact in complex ways, two of which commentators have found particularly puzzling. The first is that the branching structure inherits the effective or FAPP (`for all practical purposes',  \citet{bell1990against}) nature of decoherence, making for fuzzy macroscopic worlds (criticised by \citet{maudlin2019philosophy}). The second is the claim that branching does not happen to the world as a whole, but rather regionally in accordance with a dynamics in which branching propagates via physical interactions, as sketched in his \S8-6-7. This paper presents a model of branching by decoherence that separates these two features: the branches are exact not fuzzy, but the dynamics of branching is regional, and proceeds by local interactions governed by unitary time evolution. The model significantly clarifies a number of aspects of the process by which branches form, and shows how different features of the dynamics of branching can come apart: though we explicitly demonstrate its essential agreement with Wallace's account. While we aim to make (or at least make explicit) a number of original points about branching by decoherence, we have also written this essay with those (including students) encountering the Everett interpretation for the first time in mind. We hope that the simplicity of the model, and our elucidation of it, will lead to a better understanding of the relevant physics, and consequently of debates surrounding many worlds interpretations.

What follows, as is standard in decoherence theory, assumes unitary evolution \emph{un}interrupted by quantum jumps, and so holds for any foundational approach to QM -- including Everett's -- that assumes such dynamics. Thus it applies to Bohmian (\citet{durrBohm2009}) and modal (\citet{bacciagaluppiModal1996}) theories, but not to collapse (\citet{ghirardi1986unified}) theories (except for small systems and short times, if the probability of collapse depends on the number of degrees of freedom). Indeed, we stress that this paper concerns decoherence in general, and does not take any (further) view about the correctness of Wallace's interpretation. To put the point another way: we explain how \emph{entanglement} by unitary evolution can give rise to \emph{decoherence}, and how the latter in turn can give rise to \emph{branching} in a sense that we explicate, but we do not take up the further question of whether branches can be thought of as different `\emph{worlds}' of experience, with all the attendant philosophical questions of emergence, identity, probability and so on. We hope that distinguishing the concepts of entanglement, decoherence, branching from each other (and from that of worlds), and explicating their relationships in a more interpretation-neutral way, will be fruitful for discussions of interpretations of QT.\footnote{\citet{bacciagaluppidecoh2020} also approaches decoherence in an interpretationally neutral way in this review article, allowing for the discussion to apply across different interpretations. While we focus mainly on Everett here, as we have stated there is still much of our views on entanglement and decoherence that is relevant across the board.}

The next section (\S\ref{sec:Dec}) reviews some standard features of decoherence, especially its relationship to entanglement; although these may be known to readers, having their conceptual relations clearly in mind will help the later discussion. \S\ref{sec:simple} presents the simplest version of our model, in which an initially unentangled system becomes increasingly entangled with its environment and consequently decohered, and explains which features of Wallace's full account (here taken to be the canonical approach to branching) it does and does not instantiate. \S\ref{sec:worldmaking} then discusses how such decoherence in turn leads to local branching. Finally, the idea that branching happens regionally and locally may seem in tension with the fact of entanglement in the EPR-Bohm experiment, because, if Alice's and Bob's worlds branch independently when they separately measure the spin in the $x$-direction of their particles, it may seem mysterious that if they interact in the future and their branches join up they never find themselves in a world in which they both measured up. This issue is resolved when the model is extended to cover an initially entangled state (\S\ref{sec:EPRB}) because it is made explicit how the regional and local nature of the branching process takes place within the unitary evolution of a non-separable global state.

\section{Decoherence}\label{sec:Dec}

Quantum decoherence as a concrete material process is the suppression of interference effects of a quantum system through interaction with its environment, as in standard examples of measurement and environmental decoherence (\citet{blanchard2000decoherence}, \citet{Joosenvirondecohere1985}, \citet{joos2013decoherence}), or because part of the system acts as the environment of the rest (\citet{hackermuller2004decoherence}, \citet{lombardo1996coarse}, \citet{zeh1986emergence}). In other words, the system's state becomes entangled with that of the environment so that superposed terms in the former are correlated with those of the latter. In the formal, mathematical representation of such a process, the result is that the off-diagonal elements of the `reduced density matrix' for the system alone rapidly become FAPP negligible; where the reduced density matrix comes from ‘tracing out’ the environmental degrees of freedom. Hence the state for the system becomes approximately mixed, while the joint state itself remains pure, with interference terms remaining in the state of the joint system+environment composite. In this formal sense, \emph{any} system exhibits `decoherence' when it becomes entangled with another, its `environment', in this way. 

But we want to understand decoherence as a physical process, so what is the material significance of taking the partial trace? Moreover, entanglement is the generic outcome of unitary interactions between systems, and so in the formal sense `decoherence' is ubiquitous; but of course quantum interference is also everywhere, for instance in entangled systems such as Bell pairs -- not every concrete case of entanglement is really a case of decoherence. It must be the case then that decoherence only occurs when one of the entangled subsystems has some special features that make it count as the `environment'. But what features?

To answer these questions, let us first step back and consider a system with a single degree of freedom, $x$, though it need not be position (or even continuously valued). Suppose that the state is a superposition of states whose support is in regions $A$ and $B$: $\Psi(x)=\psi_A(x)+\psi_B(x)$. If $A$ and $B$ overlap then there will be quantum interference between the two packets: for $x$ in the overlap, (generally) $|\psi_A(x)+\psi_B(x)|^2\neq|\psi_A(x)|^2+|\psi_B(x)|^2$. But if the regions are disjoint, there will be no interference, and the probabilities for finding the system in one or the other are just as if it were classically randomly distributed between them. Now suppose a second system, with degree of freedom $y$, whose state is a superposition of states whose support lies in regions $C$ and $D$: $\Phi(y)=\phi_C(y)+\phi_D(y)$. And further suppose that the two systems become entangled through mutual interaction so that the joint state is $\psi_A(x)\phi_C(y)+\psi_B(x)\phi_D(y)$. Obviously this model massively idealizes real material systems, but its features capture some of their important general properties.

Suppose, for instance, that $A$ and $B$ mutually overlap -- the condition for quantum interference in the unentangled case -- but $C\cap D=\emptyset$. In this case, even if $x\in A\cap B$ there is no $y$ such that $\psi_A(x)\phi_C(y)$ and $\psi_B(x)\phi_D(y)$ are both non-zero; so there is now no quantum interference between the two terms, despite the overlap of the $x$ wavefunctions. As before, position of the first system is as if it were randomly distributed between $A$ and $B$, but now because of entanglement, and not because $A\cap B=\emptyset$  -- the subsystem has decohered (into an exact mixture). Graphically, $\psi_A(x)$ and $\psi_B(x)$ miss each other because the joint wavefunction moves in the $X\times Y$ configuration space, and the packets do not cross in the $y$-direction. What if $C$ and $D$ overlap as well? Then the extent to which the first system behaves classically is the extent of that overlap (i.e., to the extent that they fail to be orthogonal): if the overlap is sufficiently small, then the system's probabilities are classical, not exactly, but FAPP. 

With this general picture of decoherence in mind, let us return to the questions of the meaning of the partial trace, and what beyond entanglement (and approximate orthogonality of the environment states) is necessary for decoherence.

First, taking the partial trace of a bipartite system formally means projecting the joint state down to a state for one part, weighted according to the state of the other (as explained in standard textbooks). For instance, in our example the result is a mixture of $\psi_A(x)$ and $\psi_B(x)$, weighted by $\int|\phi_C(y)|^2\mathrm{d}y$ and $\int|\phi_D(y)|^2\mathrm{d}y$, respectively. Such a prescription makes sense when one has no knowledge of the second system beyond the joint wavefunction. Indeed, that is the essential idea of the partial trace: the joint and reduced states entail exactly the same expected values for all observables of the remaining part alone. That is, the partial trace captures everything about the state of a subsystem that the joint state does, when we don't get any more knowledge about the rest of the system from measurement, which of course is the only way to obtain any such knowledge. Formally then, taking the partial trace gives a redescription discarding information about one part of a system, while retaining all the information about another. 

This understanding of the partial trace points to an answer to our second question, regarding the relationship between entanglement and decoherence. Decoherence, after all, involves appropriate entanglement \emph{and} taking the partial trace, but when is it `appropriate' to take the partial trace and redescribe a concrete system? Not for just any joint system, but only when one knows the outcomes of measurements on one part but not the other; the former we can now generally call the `system', and the latter its `environment'. The ignorance involved is not a merely formal property of the system, since it relates to a concrete `knower' with certain capacities and limitations for knowledge. But since this answer is philosophically unsatisfactory because it relies on the concept of a `knower': Just whose ignorance? Why them? And why is it a subjective notion, given that it is relevant to physical features?\footnote{Similar questions arise for the Boltzmann entropy in statistical mechanics; the answers are also parallel, as \citet[\S2.1]{myrvold2022philosophical} stresses.} Well, consider how such ignorance might arise. In some way the environment is inaccessible: one might lack the apparatus to perform the measurement, perhaps because the relevant degrees of freedom are too small, or too numerous; or parts of the environment might somehow be physically inaccessible, perhaps because entanglement spreads so rapidly; or perhaps we simply choose to ignore the environment; and so on. (The first two cases are exemplified when a charged system interacts with the electromagnetic field since it has degrees of freedom at every point and entanglement propagates at light speed, and so they help explain the ubiquity of decoherence.) In other words, we should not think of tracing out the environment as simply relative to an agent, but instead as \emph{relative to the measurements available} to the agent; as relative, that is, to a specific `observer'. And so decoherence should be understood as (appropriate) entanglement \emph{plus} a restriction on the available interactions with the environment, both concrete material conditions (and not subjective in any troubling sense).

Now we are in a position to address a response to the example given earlier in this section; so doing will slightly generalize our understanding of decoherence. If one can manipulate the second system to move the two packets in the $y$-direction so that they come to overlap, then of course in the overlap there will again be quantum interference; and if they overlap substantially such interference may become observable. Thus, when such control of the environment is possible, it is no longer appropriate to take the partial trace and (re)describe system one as a classical mixture, for the subsequent evolution can depend on its actual quantum nature: it has not really decohered at all.\footnote{In these conditions one can speak of `false decoherence' \citep{belenchia2018quantum}. It is to distinguish `true' from false decoherence that we do not simply define decoherence to occur whenever a system becomes entangled.} Thus we should say that an entangled subsystem -- the `system' -- has decohered for agents when they lack the physical ability to observe or manipulate the remainder of the joint system -- the `environment' -- in a sufficiently fine-grained way. (Though it is worth emphasizing that in many many cases, realistic agents have effectively no control whatsoever of the environment, so the relation is moot and decoherence just is entanglement.)

To put the point even more objectively: decoherence is for us a physical \emph{relation} between two parts of a joint system \emph{and} a concrete set of physical operations.\footnote{We should acknowledge that Wallace's account of decoherence \citealp[\S3.5-10]{wallace2012emergent} does not appeal to operational limitations in this way; he understands it simply in terms of non-interfering classical histories (see also \citet{saunders21} for another, recent view that does invoke operational limitations).  Our approach is easier to explain, and equivalent, at least in the simpler model.} Of course, this relativity means that any indeterminacy regarding which operations, implies indeterminacy in the question of whether a system has decohered (and hence branched); an indeterminacy in addition to the question of just what counts as FAPP, though of a similar kind.

\section{A simple model of branching}\label{sec:simple}

Having explained the relationship between entanglement and decoherence, this section and the next turn to the question of how decoherence produces \emph{branches}, and the difference between the concepts of branching and decoherence. (We emphasize again that we remain neutral on the further question of whether branches should be thought of as `worlds' \'a la Everett). \citet{wallace2012emergent} begins his discussion with the observation that the standard account of decoherence theory makes “[…] it somewhat difficult to appreciate how exactly it is that the quantum state has the structure of a collection of quasi-classical branching worlds. We may have established that the density operator of such systems is diagonalized in a quasi-classical basis, but it is not immediately obvious how to read the branching structure off from this observation.” (p.85-6).\footnote{Two comments: (1) Wallace uses `world' where we use `branch', but for him the terms are interchangeable. (2) The notion of quasi-classicality is required for a full and realistic account of branching, since unitary decoherence is only FAPP for real physical systems.}

The model that we introduce to address these issues is based on the idealisation of discrete subsystems and interactions, making decoherence exact. It can be viewed as a simplification of the Coleman-Hepp-Bell model.\footnote{\citet{hepp72}, \cite{bell75}. Others who have considered similar models more recently include \citet{zurekEnviron1982}, \citet{CucchiettiBath2005},  \citet{DawsonSpinbath05}, and \citet{wallace2010quantum}.} The model shows that the essentials of branch propagation through an environment by entanglement can be represented without the conditions required for a more realistic treatment. Hence, some important features of Wallace's overall account that are necessary for a more realistic treatment are not features of the model. However, the model captures central features of branching by decoherence, and the massive gain in simplicity makes them so transparent as to make the trade-off worthwhile.

The model realizes the decoherence of a system of interest as it becomes entangled with its immediate environment; and the decoherence of degrees of freedom in the environment, because of their entanglement with other environmental degrees of freedom. More specifically, since we are interested in understanding the formation of decoherent branches in accordance with realistic physics in salient respects, we assume a spatially localized system, and a spatially extended environment -- a field -- with local interactions, and a finite speed of propagation. For formal and notational simplicity and transparency we pick the simplest possible quantum system (a qubit), and reduce the infinity of field degrees of freedom to a single quantum degree of freedom per unit volume (or interval, in one dimension).

Bearing in mind the discussion of \S\ref{sec:Dec} of the system-environment split, let the system be a qubit with state $|\psi\>$ lying in the Hilbert space $\mathcal{H}_s$ with (orthonormal) basis $\{\1,\0\}$; and let the environment (the `field') be a collection of spin-$\frac{1}{2}$ systems each located at a 1-dimensional lattice point $i\in\{\dots,-1,1,2,3,\dots\}$, with state $|\sigma^i\>$ lying in the Hilbert space $\mathcal{H}_\sigma^i$ with basis $\{\u,\d\}$. The system is located at the point 0.\footnote{\label{ftnt:loc}We have to make a choice about whether the field can be co-located with the system: i.e., there could have been a spin at $0$, and interactions between the system and that spin. Our choice makes no formal difference, but it allows us to identify degrees of freedom, hence branching, with regions of space. One should bear in mind, however, that in reality multiple degrees of freedom are co-located, so that what we say below about branching of a region due to entanglement with another region will also apply to branching of degrees of freedom due to entanglement with co-located degrees of freedom. Note also that if the lattice is infinite, then the field Hilbert space is $[\mathbb{R}]$-dimensional ($[\mathbb{R}]$ the cardinality of the reals), and hence non-separable. For simplicity, we assume that the lattice is bounded in both directions, even if we do not explicitly indicate this in our notation.}

We specify the following unitary interaction $U_{si}$ between the system and adjacent spin element of the environment, acting on $\mathcal{H}_s\x\mathcal{H}_\sigma^i$:
\begin{equation}\label{eq:s-eint}
\begin{split}
    \0\x\u \to \0\x\d \qquad&\qquad \0\x\d \to \0\x\u \\
    \1\x\u \to \1\x\u \qquad&\qquad \1\x\d \to \1\x\d;
\end{split}
\end{equation}
while the interaction $U_{ij}\in U[\mathcal{H}_\sigma^i\x\mathcal{H}_\sigma^j]$ between any two (adjacent) spin elements is given by:\footnote{\label{fn:non-symU}The reader may observe that the interactions are not symmetric under spin reversal (they are `controlled-not' operations in quantum computation); they are tailored to a specific asymmetric initial state (and unitary). We switch to a symmetric evolution below.}

\begin{equation}
\label{eq:e-eint}
\begin{split}
    \u\x\u \to \u\x\u \qquad&\qquad \u\x\d \to \u\x\d \\
    \d\x\u \to \d\x\d \qquad&\qquad \d\x\d \to \d\x\u.
\end{split}
\end{equation}

These interactions allow us to model how the qubit entangles locally with spin-field degrees of freedom, and how that entanglement propagates locally through the field, as in the following example: suppose that the system is located at $0$, and in a coherent superposition of states, $\0+\1$ (ignore normalization unless relevant); suppose that the field only extends in the positive half-line, with all spins initially in the state $\u$. Then at $t=0$ the state can be written:
\begin{equation}
    |\psi\>\x|\sigma^1\>\x|\sigma^2\>\x|\sigma^3\>\x|\sigma^4\>\x\dots = \b(\0+\1\b)\x\u\x\u\x\u\x\u\x\dots.
\end{equation}
Let time be discrete like space, and consider the propagation of entanglement by sequential application of the interactions defined in (\ref{eq:s-eint}-\ref{eq:e-eint}).\footnote{\label{fn:TDep}Contrary to the usual assumption, this dynamics is time-dependent, since spin-spin interactions are not constant, but turn on and off. This amounts to a sort of non-locality, but not one that undermines the relevant locality of the model. Again, we have made this choice for simplicity (and unitarity).}

\begin{equation}\label{eq:model1}
    \begin{split}
        t=0: \qquad\qquad  &\b(\0+\1\b)\x\u\x\u\x\u\x\u\x\dots\\
        \ \\
        &\qquad\qquad U_{s1}\x I\x I\x I\x\dots\\
      &\qquad\qquad\qquad\qquad\downarrow\\
                     \ \\
  t=1: \qquad\qquad  &\b(\0\x\d+\1\x\u\b)\x\u\x\u\x\u\x\dots\\
                      \ \\
&\qquad\qquad I\x U_{12}\x I\x I\x\dots\\
        &\qquad\qquad\qquad\qquad\downarrow\\
                    \ \\
   t=2: \qquad\qquad &\b(\0\x\d\x\d+\1\x\u\x\u\b)\x\u\x\u\x\dots\\
                \ \\
        &\qquad\qquad I\x I\x U_{23}\x I\x\dots\\
        &\qquad\qquad\qquad\qquad\downarrow\\
                     \ \\
  t=3: \qquad\qquad &\b(\0\x\d\x\d\x\d+\1\x\u\x\u\x\u\b)\x\u\x\dots\\
                     \ \\
  &\qquad\qquad\qquad\mathrm{and\ so\ on.}
    \end{split}
\end{equation}

The remainder of the section explains how the model exemplifies (or not)  key features of branching by decoherence. First, in the model:\\

\WallComp entanglement of the system with the environment can readily be seen to propagate in space over time. \\

Moreover, if we take (each degree of freedom of) the field to be an `environment' in the sense of \S\ref{sec:Dec}, so that it is physically inaccessible, then we are justified in tracing out the field degrees of freedom, and decoherence occurs for the system as soon as it becomes entangled with the environment. At $t=0$, tracing over the environment leaves the system in a pure state, while at later times it is in a mixture of $\0$ and $\1$ states. For instance, at $t=1$ the qubit density matrix is of course

\begin{equation}\label{eq:DM}
    \begin{pmatrix}
    1/2 & 0 \\
    0 & 1/2
    \end{pmatrix},
\end{equation}
an even classical mixture of $\0$ and $\1$ states. Such diagonalisation as systems interact makes for branching.  As Wallace says:\\
    
    \WallComp  “‘branching’ (relative to a given basis) is just the absence of interference [in a subsystem].” (p.88).\\

Now, in the terms of the previous section, the absence of interference is decoherence, not branching. As the next section explains, we take branching to be conceptually richer than decoherence as (approximate) diagonalization; and the latter is only a sufficient condition for branching. For now simply note that at each time an additional spin becomes correlated with the pre-branching state of the qubit -- sometimes also described as a `record' of the `measurement' of the spin by the field. By agreeing on the initial spin, all the correlated terms `belong together' (while anti-correlated terms `belong apart'), and for this reason it is thus such correlated values that make a branch. As we discuss further below, of course then the model shows how two branches are created and grow over time. 

Next, crucially, in the model, branching is local.\\

\WallComp 	\label{WC:local}Decoherence is a local process, so branching is a local process.\\

This feature is captured because the only interactions are between adjacent degrees of freedom, giving a discrete model of the local interactions of continuous fields.  As Wallace says:

\begin{quote}
    “[Branching...]propagates outwards at the speed of whatever dynamical interaction is causing decoherence—in practice, it propagates out at the speed of light.” (p.307)
\end{quote}

Thus the example shows explicitly how entanglement, decoherence, and branching propagate locally. The \emph{central} advantage of this model (justifying its idealizations) is that makes this locality explicit.\\

\WallComp 	Branching is with respect to a basis because the reduced density matrix will in general not diagonalize in all bases.\\

To see how, we would need to change the initial choice of qubit state, because the density matrix (\ref{eq:DM}) has the special property of being the same in all bases.\footnote{For any unitary operator, $\hat U^\dag\frac{1}{2}I\hat U = \frac{1}{2}\hat U^\dag \hat U = \frac{1}{2}I$.} But if instead the initial state had been $\sqrt{\frac{2}{3}}\0+\sqrt{\frac{1}{3}}\1$, say, then the reduced density matrix at $t=1$ would be, in this basis, the mixture

\begin{equation}
    \begin{pmatrix}
        2/3 & 0\\
        0 & 1/3
    \end{pmatrix}.
\end{equation}
But in, say, the $\{\sqrt{\frac{1}{2}}(\0+\1),\sqrt{\frac{1}{2}}(\0-\1)\}$ basis it becomes,

\begin{equation}
    \begin{pmatrix}
        1/2 & 1/6\\
        1/6 & 1/2
    \end{pmatrix},
\end{equation}
so it is no longer diagonal. Of course, that a state is a mixture is a basis independent fact; what the failure to diagonalize means is that the qubit is not a mixture of the new basis states, but a mixture of their quantum superpositions. 
Or, in concrete physical terms, it is in this sense that branches are basis-relative: when a system decoheres, it has a special set of physical properties, singled out by the fact that they (and typically, they alone) are distributed with classical probabilities.\\

\WallComp 	Decoherence produces branches that are effectively dynamically independent of each other, and that behave quasi-classically.

\begin{quote}
    “\dots if the quantum state is a faithful description of physical reality, then that physical reality consists of a vast number of distinct [branches,] dynamically speaking almost independent of one another; by and large they behave approximately classically...” (p.46)
\end{quote}

Our branches are dynamically independent: for instance, the state of the qubit in one branch evolves independently of the state in the other (indeed, it remains constant once the branch forms), and of course there is no interference between the two branches. However, given the linearity of the underlying unitary dynamics, these properties trivially represent the exact orthogonality of the states in our idealization. The model does not illustrate realistic classical physics: for classical particle branches, one would need instead to have branches containing narrowly localized wavepackets (e.g., \citet[\S3.5]{wallace2012emergent}). So, the branches have exactly `classical dynamics' in the sense of possessing determinate values for the qubit (and entangled field degrees of freedom) though this dynamics is not realistic.

Now, for Wallace (\S3.5-10) a branch is found when packets in a superposition can be described by quasi-classical physics, and by feature (v) there are indeed two qubit branches in this sense. Therefore it is important that decoherence and branching is not merely a matter of the density matrix diagonalizing in a basis (signaling the disappearance of quantum interference in that basis), but also of remaining diagonalized in that basis with some definite (approximately) classical dynamics \emph{over time}.\footnote{There can of course be no continuous quantum evolution that keeps a system in the vectors of an orthogonal basis, so the basis in question must be overcomplete, as of course a set of approximately localized wavepackets is (cf.\citet[\S3.2]{wallace2012emergent}).} In reproducing emergent classical dynamics, albeit of a trivial kind, the model respects this understanding.

As mentioned previously, there are also certain significant differences between the model and a more realistic account of branching, which are explained in the rest of this section. In the model, diagonalisation is exact and not approximate, i.e. interference terms are zero in our chosen basis. This is because the model's discrete dynamics entails that the system's degrees of freedom entangle (if at all) \emph{immediately} with \emph{orthogonal} states of the environment, so that there is an exactly impure state when the environment is traced over. Thus, the model stipulates away two of the most important questions in the study of real-world decoherence: into what -- preferred, classical -- basis does the system decohere? And how quickly does it decohere into that basis? (The answers to these questions are encoded into the calculation of the `decoherence functional', e.g., \cite{DoWHal92}.) In the model the answers are simply: for the qubit the 0-1 basis and for the field the up-down basis, and after interacting for one discrete unit of time. Note, however, that in our model and in more realistic cases these matters are settled in the same way, namely by the interaction dynamics. 

Since the model does \emph{not} represent a continuous transition of a pure state into a state ever closer to an impure mixture, it does not display the following feature:\\

\WallComp 	Decoherence is understood as an approximate process that leads to an irreducible indeterminacy in branch identity, individuality and numerosity.\\

Or as Wallace puts it:

\begin{quote}
“Decoherence causes the Universe to develop an emergent branching structure. The existence of this branching is a robust (albeit emergent) feature of reality; so is the mod-squared amplitude for any macroscopically described history. But there is no non-arbitrary decomposition of macroscopically-described histories into ‘finest-grained’ histories, and no non-arbitrary way of counting those histories.” (p.101-2)
\end{quote}
In this regard our model is unrealistic: since decoherence is exact, entanglement produces exactly two branches of correlated values. In reality, the branching structure would be extremely complex with branches continually propagating into every spacetime region from every direction. However, making this idealization helps us to clarify other features of branching by decoherence, as we have started to show. Moreover, having a clear model of branch propagation allows one to study concretely branch `collisions' (albeit in idealization) and the more complex structures they produce, as we start to show in the following sections.

Note, however, that another potential source of branch indeterminacy remains: as explained in \S\ref{sec:Dec}, decoherence (in the material sense) is a relation between a joint system and a set of physical operations that distinguish system from environment. Since `$x$ decoheres' is a relation not a monadic property, then the same applies to `$x$ is a branch'; without specification of a set of operations, where the branches are, and how many there are remains indeterminate (as does the split between system and environment.) 

With this understanding of decoherence in mind, it is worth emphasizing that the central point that Bell (\citeyear{bell75}) made about the Coleman-Hepp-Bell model holds for our model too (as it will for any case of decoherence). There always exist observables sensitive to the interference between the branches, that will distinguish the entangled pure state from a mixture. But as we made clear at the end of \S\ref{sec:intro}, a system does not classicize simpliciter; rather it becomes classical \textit{relative} to the available operations. Here we are making the physically reasonable assumption that observables capable of measuring the correlations between many degrees of freedom are not available, so that interference is indeed inaccessible FAPP.\\

\WallComp 	Next, ``branching is caused by any process which magnifies microscopic superpositions up to the level where decoherence kicks in \dots'' (p.99)\footnote{\citet[99]{wallace2012emergent} outlines the three kinds of processes that cause branching involving this amplification: human experiments, 'natural' quantum measurements and chaotic systems. The details are not relevant here beyond them encapsulating the change from micro to macro. \citet{bacciagaluppidecoh2020} also emphasises the change from micro to macro as a defining feature of decoherence.}\\

This is not a feature of the model, in the sense that decoherence kicks in immediately, not in a continuous way; therefore, even if more realistic physics requires such smooth amplification, in principle it is sufficient but not necessary for branching. However, the model does show the propagation of entanglement through space, from a single qubit to many entangled spins, so it does involve the decoherence of a `microscopic' system by a `macroscopic' one. Moreover, while we count a single discrete field degree of freedom as an environment, this is of course because it idealizes the many field degrees of freedom associated with a region of space: the immediate, exact decoherence of the model idealizes the continuous, approximative decoherence that characterizes Wallace's account. Moreover, it is similarly incapable of showing that, according to the Born Rule, the overwhelmingly probable branches are definite in the right classical properties obeying the right classical dynamics. 

Related to this point is the following feature of decoherence. Realistic semi-classical system are subject to opposing tendencies: quantum dispersion due to non-linear self-interaction and continual decoherence with the environment suppressing interference -- the tendencies for the wavefunction to spread, and that spread to have no effect. Of course, the result is that new branches form, but with probabilistic weights that start small and grow over time. Our very simple discrete dynamics is far too simple to model this kind of behaviour at all. \\

\WallComp The environment carries a record of the \emph{history} of the entangled system. \\

Because a real system is constantly interacting with its environment, its state is constantly being `measured' by the field. (Moreover, because the parts of a realistic environment are constantly in interaction with one another, `news' about the results of these measurements are constantly propagating through the environment.) The state of the environment (and indeed of parts of it) at any time is correlated with all the `news' it has so far received, and hence with the history of the system. Since our model involves a one-off interaction rather than a series of environmental measurements of the system, the modelled environment only records the initial measurement result, not a non-trivial history.

We end this section with a feature that our model does replicate.\\

\WallComp 	the structure of the wavefunction ``is not fundamentally tree-like... The branches are not fundamental, but they are not any less real for this restriction.” (p.63)\\

Branches are not a basic posit in Wallace's account, they arise (if at all) in a system from the structure of its state space, and its dynamics. Contrast this, for instance with a naive approach to many worlds quantum theory, in which measurement causes a spontaneous division of the whole world into many new worlds. Our simple model demonstrates (subject to the limitations we have described) that branches are created by the standard elements of quantum mechanics, not by positing `worlds' and `splitting' as new fundamental elements. However, the comparison to the splitting of whole worlds raises the question of whether classical branches \emph{extended in space} occur in our model, and how. They do, as we explain in the next section.

\section{Ways of branch-making}\label{sec:worldmaking}

In our example, the interactions mean that entanglement propagates outwards from the qubit located at $0$, one degree of freedom at a time, taking a finite time to reach any given part of the field. Hence, in the model:\\

\WallComp 	Branching is not (necessarily) global.\\

This is stated explicitly by Wallace:

\begin{quote}
    “As we would expect from the absence of action at a distance, then, branching is not a global phenomenon. Rather, when some microscopic superposition is magnified up to macroscopic scales (by quantum measurement or by natural processes) it leads to a branching event which propagates outwards at the speed of whatever dynamical interaction is causing decoherence—in practice, it propagates out at the speed of light.” (p.307)
\end{quote}

To explain, we need to clarify what is meant by `branch' further, as there is a certain ambiguity. Up to this point we have primarily considered the two states (in the diagonal basis) of the decohered subsystem as branches, and in many cases one is indeed only interested in the `regional branching' of a given subsystem in this way. For instance, to explain the apparent `collapse' of the pointer state of a measurement device after an experiment, in terms of its branching due to the entanglement of the pointer with its typically macroscopic environment, including the air, electromagnetic field, sensory organs of the experimenter, etc, etc. Let's call this kind of branch a `subsystem branch'.\footnote{We don't say `local branches', since we use `local branching' to refer to the \emph{process} of decoherence, which can produce both subsystem and `extended' branches, as we shall now explain. Moreover, we will see later that subsystem branches can have spatially disconnected parts.}

But, like Wallace above, one can also think of a branch as extended, and comprised of many subsystems: not just the pointer, but everything with which it is entangled. For instance, in, say, the state $\b(\0\x\d\x\d\x\d+\1\x\u\x\u\x\u\b)\x\u\x\dots$ at $t=3$, there are not only two subsystem branches at 0, but also at each location $1-3$, in $u$ or $d$ field states; the field degrees of freedom at these locations are entangled with each other, and in the model each is an `environment' in the defined sense, so each exhibits decoherence. Moreover, the $\1$ branch at $0$, for instance, forms a larger extended branch with the $\u$ branches located at $1-3$ -- and not with the $\d$ branches. 

We can use our model to understand how subsystem branches combine to form such `extended branches', by appeal to the \emph{total}, unreduced, wavefunction (at $t=3$), and the \emph{correlations} that it encodes between the qubit and the spins: $\1$ appears only in the same term as $\u$ for any of the field degrees of freedom $1-3$, representing the (material) fact there is a 100\% Born Rule probability of finding them together. It is for this reason that the $\1$ branch at $0$ forms an extended branch with the $\u$ branches located at $1-3$ -- and not with the $\d$ branches. Of course, correlation is not sufficient for forming an extended branch, for it is not the case that just any entangled subsystems form a branch, since they may still behave quantum mechanically, not classically. As we have discussed, classical behaviour requires decoherence. So more carefully, a branch in the extended sense is comprised of entangled subsystems, each of which is decohered; which, as we have emphasized, means that each is entangled with an uncontrollable environment (in this case the other subsystems) so that it exhibits no quantum interference.\footnote{If the field is infinite, so at no time does a `global branch' form; but it could if the field is finite, because in a compact dimension, say.}

Because of these correlations, every entangled part of the field `knows' the state of the qubit, because the result of its `measurement' by the spin at 1 has propagated through space.\\

\WallComp The degrees of freedom that compose an extended branch agree about the classical state arising from the initial decoherence event.

\begin{quote}
    ``the information about the system recorded in the original decoherence process will be distributed very widely across the environment''. (p.89)
\end{quote}

That they all agree with certainty is an artifact of our model; more realistically some environmental degrees of freedom may lose the information, and the probability of agreement will only be approximately 100\%. But of course it is this agreement that ultimately makes for an extended branch at all, for what else do the various states attributed to the various degrees of freedom have to do with one another? 

And so just as decoherence is suitable entanglement plus a lack of control of the environment, so branching is decoherence plus systematic correlations between the entangled degrees of freedom. (Of course, such correlations follow from decoherence, so the difference is only conceptual.) Now, the correlations rely on the Born Rule, so concretely they amount to probabilities that the parts of the branch be `found' in certain states and not others. This statement -- and hence the material, physical conception of a branch -- is inadequate, but as far as we can go without taking a stance on the interpretation of QT. For, what does it mean to `find' a system in a state? Fully answering such a question requires one to take stance on the physical significance of the wavefunction and its relation to measurement outcomes, and so taking a stance on the interpretation. Since our goal is to provide an account of branching that is relatively \emph{neutral} on interpretation, we leave these questions incompletely unanswered here.\footnote{There is a second issue in cashing out extended branches in terms of `finding' degrees of freedom to be correlated, since by definition the relevant environmental degrees of freedom cannot be (appropriately) measured! So the concept of an extended branch involves a hidden counterfactual about different possibilities of interacting.} 

Nonetheless, our interpretation-neutral conception of branches is still rich enough to be conceptually useful, because even interpretations that differ on the nature of measurement agree that the Born Rule gives probabilities for measurement outcomes. Whether and what branches amount to beyond that depends on the interpretation: for a branch to be a world requires not only that its subsystem values be `found' to be correlated if inspected, but to be \emph{realized}. If \emph{all} the values (in the diagonal basis) of each subsystem are so realized, then the many branches are many worlds. If only the values in \emph{one} branch are realized -- because, say, a hidden variable or modal state selects that branch -- then of course that is the only branch. Seeing things this way illuminates the value of modelling and clarifying a fairly interpretation-neutral concept of a branch: first to help understand (and hence debate) the specific commitments of a many worlds interpretation; second because the process of branching occurs according to any version of QT that involves unitary evolution, and so must be taken into account in applying any such interpretation. 

We end this section, however, by emphasizing that the model exemplifies how we have given up the idea that the whole world branches instantaneously on measurement; rather it splits into extended branches that grow over time. Indeed, the central aim of this paper is to explore this process and how the world is divided into distinct branches across spacetime, and to that end we consider various modifications of the model in what follows. 

First, consider the following: suppose, continuing on from (\ref{eq:model1}), that between $t=3$ and $t=4$ the qubit interacts with a spin at $-1$:
\begin{equation}
    \begin{split}
       \u\x\b(\0\x\d\x\d\x\d+\1\x\u\x\u\x\u\b)\x\u\x\dots \to\\ \b(\d\x\0\x\d\x\d\x\d+\u\x\1\x\u\x\u\x\u\b)\x\u\x\dots.
    \end{split}
 \end{equation}
That is, the interaction  that correlates the qubit with another degree of freedom, will typically also bring that degree of freedom into correlation with the entangled spin at $3$; `measuring' the qubit also `measures' the entangled spin. But crucially, this is not a property of the state alone, but also of the kind of interactions that occur in nature; that there is local interaction between the qubit and adjacent spins that leaves distant spins unchanged. If this were not the case, and correlation between the qubit and distant spin were typically destroyed by measurements of the qubit, there would be little practical reason to view their branches as part of a larger correlated whole. Sewing together distant branches to form an extended branch (and so, perhaps, a world) is then not just a matter of the form of the state, but also of dynamical stability of correlations.

\section{When branches collide}\label{sec:collide}

We can also modify the model to study decoherence and the growth of branches in more interesting cases: those with a \emph{pair} of systems, entangled or not, in an environment. So first let there be two separable qubits, both initially in the state $\0+\1$, located at $0$ and $5$; and let the spins all be in the state $\u$. The qubit-spin interaction is as before, but modify the field self-interaction $U_{ij}$:

\begin{equation}\label{eq:NEWe-eint}
\begin{split}
    \u\x\u \to \u\x\u \qquad&\qquad \u\x\d \to \d\x\u \\
    \d\x\u \to \u\x\d \qquad&\qquad \d\x\d \to \d\x\d.
\end{split}
\end{equation}
The difference from (\ref{eq:e-eint}) is that we now have left-right symmetry, because branching occurs in both directions (see footnote \ref{fn:non-symU}).\footnote{For consistency, it is also necessary to have an even number of spins.} 

The evolution of the system proceeds as follows:
\begin{equation}\label{eq:collide}
    \begin{split}
        t=0: \qquad\qquad  &\b(\0+\1\b)\x\u\x\u\x\u\x\u\x\b(\0+\1\b)\\
        \ \\
        &\qquad\qquad U_{s1}\x I\x I\x U_{\phi4}\\
      &\qquad\qquad\qquad\qquad\downarrow\\
                     \ \\
  t=1: \qquad\qquad  &\b(\0\x\d+\1\x\u\b)\x\u\x\u\x\b(\d\x\0+\u\x\1\b)\\
                      \ \\
&\qquad\qquad I\x U_{12}\x U_{34}\x I\\
        &\qquad\qquad\qquad\qquad\downarrow\\
                    \ \\
   t=2: \qquad\qquad &\b(\0\x\u\x\d+\1\x\u\x\u\b)\x\b(\d\x\u\x\0+\u\x\u\x\1\b)\\
                \ \\
        &\qquad\qquad I\x I\x U_{23}\x I\x I\\
        &\qquad\qquad\qquad\qquad\downarrow\\
                     \ \\
  t=3: \qquad\qquad &\0\x\u\x\d\x\d\x\u\x\0\ +\\
  &\0\x\u\x\u\x\d\x\u\x\1\ +\\
  &\1\x\u\x\d\x\u\x\u\x\0\ +\\
  &\1\x\u\x\u\x\u\x\u\x\1\\
                     \ \\
  &\qquad\qquad\mathrm{and\ so\ on}
    \end{split}
\end{equation}

As before, entanglement propagates through the field, though now from both directions, and where there is entanglement with the field there is decoherence and branching. At $t=0$ the qubits are unbranched, since they are unentangled with the environment. At $t=1$, there are four branches, two at $0$ and two at at $5$ (in both cases with qubit values 0 and 1) due to entanglement with the adjacent spins. Since there is no correlation between the values on one side with the values on the other (i.e., the chance of finding $0$ on one side is independent of whether $0$ is found on the other) these subsystem branches do not form an extended branch. If we were to treat the qubits as an environment for the field, then the field degrees of freedom at 1 and 4 are also branched (into up and down spin branches). Because they are correlated with the qubits, these form four extended branches: at the location $0\cup1$, a pair with states $\1\u$ and $\0\d$; and at $4\cup5$ two more with the states $\u\1$ and $\d\0$.

At $t=2$ we note that the new field dynamics erases entanglement as it is propagated: although the spins at 2 and 3 are now correlated with the qubits at 0 and 5, respectively, those at 1 and 4 are separable, and hence are no longer correlated with the qubits. In other words, the situation is just as for $t=1$, except for the field at 2 and 3, rather than 1 and 4. This loss of entanglement is unrealistic but it is a consequence of the left-right symmetry of this simple model. In reality, correlations with the qubit would remain, and there would be a pair of extended branches in each of the regions $0\cup1\cup2$ and $3\cup4\cup5$.

A similar issue applies at $t=3$: the field at 2 becomes uncorrelated with the left qubit, but correlated with the right qubit (and conversely for the field at 3). In reality of course, the field in these regions would be correlated with both qubits. We can represent this state of affairs by considering the joint state of the field degrees of freedom at $2\cup3$, which is correlated with both qubits: there is, for instance, no chance of finding the $\d\x\d$ state unless the qubits take the value 0. On this understanding (and again pretending that the field at 1 and 4 is also entangled with the qubits), now there still are four extended branches, represented by the four components of the joint state, all extending across the \emph{whole} of the lattice from 0 to 5. Did something non-local just happen as the branches went from encompassing three locations to six in a single step? In a loose sense, of course \emph{yes}: local qubit branches, for instance, instantaneously joined in extended branches faster than influences could propagate between them. But in the important sense, \emph{no}, for the dynamics is perfectly local, so all that happened is that entanglement spread locally. In particular, note that although the branches encompass both qubits, they remain uncorrelated: the chance of finding 0 on the left is independent of finding it on the right, for example. Their values find themselves in the same branch, not because they are correlated, but because they are both correlated with the field at $2\cup3$. That is, the model helps illuminate how branches will grow discontinuously when they meet, even though the dynamics, hence the propagation of information about distant events, is purely local.

\section{Non-local correlations from local branching}\label{sec:EPRB}

Finally, to further explore the issue of locality, let us model the same setup, except with initially entangled qubits, in the state $\0\x\1+\1\x\0$.
\begin{equation}
    \begin{split}
        t=0:\qquad \0\x\u\x\u\x\u\x\u\x\1 &+ \1\x\u\x\u\x\u\x\u\x\0\\
        \ \\
        U_{s1}\x I&\x I\x U_{\phi4}\\
      &\downarrow\\
                     \ \\
  t=1: \qquad  \0\x\d\x\u\x\u\x\u\x\1 &+ \1\x\u\x\u\x\u\x\d\x\0\\
                      \ \\
I\x U_{12}&\x U_{34}\x I\\
        &\downarrow\\
                    \ \\
   t=2: \qquad\0\x\u\x\d\x\u\x\u\x\1 &+ \1\x\u\x\u\x\d\x\u\x\0\\
                \ \\
        I\x I\x & U_{23}\x I\x I\\
        &\downarrow\\
                     \ \\
  t=3: \qquad\0\x\u\x\u\x\d\x\u\x\1 &+ \1\x\u\x\d\x\u\x\u\x\0\\
                     \ \\
  &\mathrm{and\ so\ on}
    \end{split}
\end{equation}

This set-up shows that:\\

\WallComp Despite the local propagation of branches, the non-separability of quantum states will affect their unfolding, and lead to Bell inequality violating records (in this case, of the measurement of the qubit by the field).
\begin{quote}
    “When two branching regions intersect, if the branchings were caused by decoherence acting on some previously entangled system whose components were at the respective branching centres, the nonlocal information associated with the entanglement propagates outward and serves to determine just how the branches intersect.” (\citet[322]{wallace2012emergent})
\end{quote}

Perhaps the most important feature of the model is that it clarifies the ways in which branching is at once local and non-local. To see how, let us follow through the example just given. Initially, although the qubits are mutually entangled, there is no decoherence (and hence no branching) if we take them as `system' not environment; and of course they are unentangled with the field. (However, if their values were not to be compared at a later time, the distant qubit would now effectively be the environment for the other, and it would be branched, in to 0 and 1 states.)

What about at $t=1$? One might at first think that, as in the previous -- unentangled -- example, there are four branches: $\0\x\d$ and $\1\x\u$ at $0\cup1$, and $\u\x\1$ and $\d\x\0$ at $4\cup5$. But that is not correct, because the values at the two regions are (anti-)correlated, so there are rather two branches, both extended over the (disjoint) region $0\cup1\cup4\cup5$: one with the values 0 and $\d$ on the left and $\u$ and 1 on the left; the other with 1 and $\u$ on the left and $\d$ and 0 on the right. 

The two branches then propagate in both directions towards the centre, until they meet at $t=3$ as before, and extend across the whole region 0-5.\footnote{Again ignoring the limitation that the correlations are lost in the model as entanglement propagates.} If one compares this state with the final one in (\ref{eq:collide}), one sees that, indeed, only two of the branches from that case are present. Of course this is no surprise, as there were only ever two of the terms present: there were no initial terms in which the qubits were in the same state, as noted at $t=1$.

One can think of $t=3$ as the moment at which correlations may be locally observed; and in a suitable experiment, measuring the field spins at multiple angles, violations of the Bell-inequalities would be found, showing that the left and right parts of the branches must have `known about' each other before they met. But it is explicit in the model that this knowledge is not propagated faster than the branches; instead it was of course always present, initially in the non-separable qubit joint state, and then in the qubits plus entangled environment state -- there were not four initial terms, but two, each correlating two (disjoint) parts. All the while, the actual propagation of the branches is completely local by construction.\footnote{Two caveats. (i) We emphasize that we only claim the locality of branching; whether the dynamics is fully local is interpretation-dependent. Since there is nothing but branching in many-worlds dynamics, then the account is fully local. Bohmian mechanics, though, also has dynamics for hidden classical variables; these have to be correlated at a distance, from the start, and so interact non-locally. But this, of course, is a key feature of the theory. Thanks to Eric Winsberg for asking about this. (ii) Even if the right qubit does not interact with the field, then at $t=1$ qubit entanglement means that the branches still occupy an extended (disconnected) region, now $0\cup1\cup5$. In that sense the branches can propagate `non-locally', from 0 to 5. This effect is of course simply a reflection of the fact that the right qubit state decohered because it became entangled with the field at 1 -- via its prior entanglement with the left qubit, so not because of any non-local interaction.} 

Given that a central concern with violations of the Bell inequalities is the implication for the relation between QT and relativity, it is worth observing that:\\

\WallComp In virtue of its locality, branching is relativistic.\\

Our model is non-relativistic, insofar as it assumes a global time.\footnote{Indeed, as noted in footnote \ref{WC:local}, the dynamics depends on this global time.}  However, insofar as it only assumes an interaction between adjacent subsystems (Feature \ref{WC:local}), it captures the key aspect of a relativistic system, namely the finite propagation speed of the entanglement producing branches. Branches would form in just the same way in a relativistic elaboration of our model: non-locality arising from initial non-separability propagating in a perfectly relativistic way.

\section{Conclusion}

The complexity of decoherence-modelled branching often leads to key features being obscured. The model here allows us to focus on only a few of these features, highlighting those that are sufficient for a local branching structure. These features, despite being outlined by Wallace, play a role that can easily be obscured by the FAPP nature of branching in a more realistic account. 

The simplicity of the model is to its advantage. By removing certain realistic features such as inexactness, we are able to clearly outline the relationship between entanglement, decoherence and branching. Decoherence is entanglement \emph{plus} a lack of control over the environment. Branching is then decoherence \emph{with} the systematic correlation between the entangled degrees of freedom. The model highlights the local propagation of branching, allowing for a distinction to be made between different ‘types’ of branches. For example, a subsystem branch is merely the decohered subsystem, which is not necessarily a conceptually rich branch. On the other hand, extended branches start with the decohered subsystem but extend to all other parts that become entangled with it. It is these extended branches that spread (locally) and the dynamics controls how distant correlations are formed, maintaining Bell-type scenarios. 

Extension of the model from a single qubit state to a pair of entangled qubits allows for a systematic analysis of certain dynamical features of decoherence and branching and gives us key insights into the mechanics of it. By not representing some features of a more realistic account of branching, the model perspicuously represents the spread of entanglement and how it leads to local branching despite nonseparability. The model is as interpretationally neutral as possible on the full nature of branching, allowing it to be applicable across quantum interpretations that take unitary evolution without collapse seriously. Importantly, these (extended) branches can be conceptually distinguished from  ‘worlds’.  As we discussed in \S\ref{sec:simple}, the difference between a branch and a world lies in whether the definite physical quantities associated with the former are in fact \emph{realized}. If not, we only have (Born Rule) probabilities for certain values to be `found'; if so, then the possession of the values is actual, and we have a world. Any `realist' interpretation of QT seems committed to the physical realization of some branch values, and the `many worlds' view is a specific form of such realism: that the values associated with \emph{every} branch are realized. Wallace (103ff) argues that, as such, it is nothing but realism about `quantum mechanics taken literally'; other realistic approaches involving modifications to the basic formalism in the form of dynamical collapses or hidden variables, for instance. We do not take a position here, but offer our discussion of branching -- which is a matter of fact for all such interpretations -- as establishing some common ground on which comparisons of such realist interpretations may be fruitfully conducted.

\bibliographystyle{plainnat}

\bibliography{bibliography}

\end{document}